\documentclass[aps,prb,floatfix,twocolumn,showpacs,superscriptaddress,groupedaddress]{revtex4-2}
\usepackage{amsmath}
\usepackage{amssymb}
\usepackage{bm}
\usepackage{hyperref}
\hypersetup{colorlinks=true,breaklinks,linkcolor=blue,urlcolor=blue,citecolor=blue}
\usepackage[dvipdfmx]{graphicx}
\usepackage{physics}
\usepackage{braket} 
\usepackage{color}

\begin{document}
  \title{Many-body Chern insulator in the Kondo lattice model on a triangular lattice}
  \author{Kota Ido}  
  \affiliation{Institute for Solid State Physics, University of Tokyo, 5-1-5 Kashiwanoha, Kashiwa, Chiba 277-8581, Japan}
  \author{Takahiro Misawa}
  \affiliation{Institute for Solid State Physics, University of Tokyo, 5-1-5 Kashiwanoha, Kashiwa, Chiba 277-8581, Japan}

\begin{abstract}
The realization of topological insulators induced by correlation effects is one of the main issues of modern condensed matter physics.
An intriguing example of the correlated topological insulators is a magnetic Chern insulator induced by a noncoplanar multiple-$Q$ magnetic order.
Although the realization of the magnetic Chern insulator has been studied in 
the classical limit of the Kondo lattice model, research on the magnetic Chern insulator in the original Kondo lattice model is limited.
Here, we investigate the possibility of the many-body Chern insulator with the noncoplanar triple-$Q$ magnetic order in the Kondo lattice model on a triangular lattice.
Using the many-variable variational Monte Carlo method, we reveal that the triple-$Q$ magnetic order becomes a ground state at quarter filling in an intermediate Kondo coupling region. We also show that the many-body Chern number is quantized to one in the triple-$Q$ magnetic ordered phase 
utilizing the polarization operators.
Our results provide a pathway for the realization of the many-body 
Chern insulator in correlated electron systems.
\end{abstract}

\maketitle
\section{Introduction}
The Kondo coupling\cite{kondo1964resistance}---magnetic interaction between localized spins and conduction electrons---has been extensively studied 
as a driving force for inducing several exotic quantum phases originally in heavy-fermion systems~\cite{Tsunetsugu1997, Stewart2001, Lohneysen2007, gegenwart2008quantum, si2010heavy}. 
The Kondo coupling leads to two main effects: the long-range Ruderman-Kittel-Kasuya-Yosida (RKKY) interactions between the localized spins~\cite{Ruderman1954,Kasuya1956,Yoshida1957}, 
and the formation of a singlet between the local spins and the conduction electrons. 
The competition between these two effects gives rise to a quantum critical point 
at which the magnetic transition temperatures become zero~\cite{DoniachPhysica1977}. 
The study of spillover effects emergent from the quantum critical points, such as non-Fermi liquid behaviors and unconventional superconductivity, 
has been one of the main topics in condensed matter physics~\cite{Stewart2001, Lohneysen2007, gegenwart2008quantum, si2010heavy}.

Moreover, the Kondo coupling has been also found to manifest 
various quantum phases, 
such as a ferromagnetic phase via the double-exchange interaction~\cite{Zener1951,deGennes1960,Kubo1972},
a charge-ordered phase~\cite{OtsukiJPSJ2009,PetersPRB2013,MisawaPRL2013}, 
a partial disordered phase~\cite{MotomePRL2010,Sato_PRL2018}, 
topological orders~\cite{Hsieh2017SA}, 
exotic superconductivity~\cite{Bodensiek_PRL2013,Ohtsuki_PRL2015} including topological superconductivity~\cite{Choi2018, Bedow_PRB2020, chang2023topological},
and a skyrmion crystal with a multiple-$Q$ magnetic order~\cite{OhgushiPRB2000,Martin2008,Kumar2010,Akagi2010,Akagi2012}. 
In particular, the emergence of a noncoplanar triple-$Q$ magnetic ordering has attracted significant interest since it can be regarded as the magnetic topological insulators, i.e., 
the magnetic Chern insulators~\cite{Chang2013,Tokura2019,Bernevig2022}, 
which show the quantum Hall effects without magnetic fields\cite{Haldane1988, Liu2008, Liu2016}.
However, previous theoretical studies were primarily 
conducted for the classical limit of the Kondo lattice models 
that ignore quantum effects. 
The quantum effects were only discussed within the linear spin-wave approximation to the ferromagnetic Kondo lattice model~\cite{AkagiJPSJ2013}.

To clarify the topological nature of the triple-$Q$ magnetic ordered state, direct calculation of the many-body Chern number is necessary. For non-interacting systems, efficient methods for calculating the Chern number~\cite{fukui2005,coh2009} have been proposed and applied to a wide range of systems. However, these methods cannot be applied to interacting systems where wavenumber is not a good quantum number. A pioneering work by Niu $et$ $al.$~\cite{Niu1985} shows that the many-body Chern number can be calculated by regarding the twisted boundary condition as an effective wavenumber. However, the numerical cost is huge in this approach, since it requires integration over all twisted boundary conditions. Recently, several efficient methods that do not require full integration have been proposed~\cite{Shiozaki2018, Kudo2019, Kang_PRL2021, Dehghani2021}. Despite this theoretical progress, it remains a significant challenge to develop an efficient and flexible numerical method that can calculate the many-body Chern number in a wide range of interacting systems with large system sizes.

In this paper, we examine the stability of the triple-$Q$ magnetic 
ordered phase in a quantum spin 1/2 Kondo lattice model on a triangular lattice by using the 
many-variable variational Monte Carlo (mVMC) method~\cite{Tahara2008a,Kurita2015,misawa2019mvmc,becca2017quantum},
which can take into account both spatial correlations and quantum fluctuations efficiently.
Our results demonstrate that the triple-$Q$ state is 
indeed realized in an intermediate Kondo coupling region at quarter filling. 
Furthermore, we show that the triple-$Q$ state has the non-trivial Chern number
by explicitly calculating the many-body Chern number 
based on the Resta's polarization operator~\cite{Resta1998,Kang_PRL2021,Dehghani2021}. 
Our findings provide a theoretical basis for understanding the emergence of magnetic topological insulators induced by the Kondo coupling.

%
\section{Model and Method}
\begin{figure}[tb]
  \begin{center}
    \includegraphics[width=80mm]{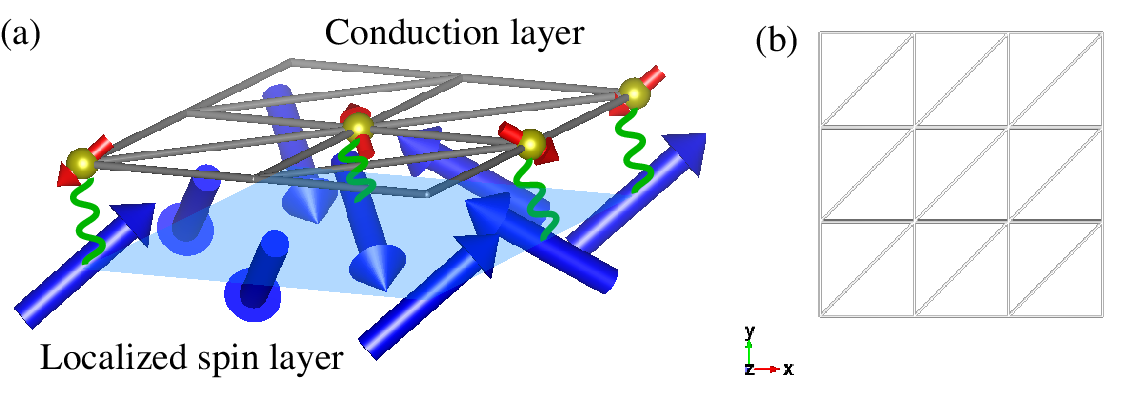}
  \end{center}
  \caption{(a) Schematic figure of the Kondo lattice model on the triangular lattice. Top and bottom layers represents conduction and localized spin layaers, respectively.
   (b) Square lattice with next-nearest hoppings deformed from the triangular lattice. 
  }
  \label{klm}
\end{figure}

%
The Kondo lattice model on a triangular lattice shown in Fig. \ref{klm}(a) is defined by 
\begin{equation}
  \mathcal{H} = -\sum_{i,j,\sigma} t_{i,j} c^\dagger_{i\sigma c} c_{j\sigma c}+ J\sum_i \bm{S}^c_{i} \cdot \bm{S}^{s}_{i},\label{eq:klm}
\end{equation}
where $c^\dagger_{i\sigma c}$ and $c_{i\sigma c}$ are the creation and annihilation operators of electrons with spin $\sigma$ at the $i$th site on the conduction layer, respectively.
$\bm{S}^{c/s}_{i}$ denotes the spin 1/2 operator at the $i$th site on the conduction($c$)/localized spin($s$) layer, namely $S^{\lambda,\alpha}_{i} = \sum_{\sigma,\sigma'} c^\dagger_{i \sigma \lambda} \sigma^{\alpha}_{\sigma,\sigma'}c_{i \sigma' \lambda}/2$ where $\bm{\sigma}^{\alpha}$ is the Pauli matrix for $\alpha=x,y,z$.
We set $t_{ij}=t$ for the nearest-neighbor sites on the lattice and $t_{ij}=0$ for the others.
We focus on the antiferromagnetic Kondo lattice system, namely $J>0$.
As shown in Fig. \ref{klm}(b), we map the triangular lattice onto the
square lattice with next-nearest hoppings. 
Both the conducting electron and the localized spin layer 
have $N_s = L \times L$ sites under the periodic-periodic boundary condition.

To analyze the Kondo lattice model on a triangular lattice at quarter filling, 
we employed the mVMC method, which enables us to obtain 
the accurate wavefunctions of ground states in strongly correlated electron systems~\cite{Tahara2008a,Kurita2015,misawa2019mvmc,becca2017quantum}.
As a trial wave function for the VMC method, 
we adopted the generalized pair product wave function 
with the Gutzwiller-Jastrow correlation factors, 
which are given as
\begin{align}
&\ket{\psi} = \mathcal{P}_G \mathcal{P}_J \ket{\phi},\\
&\ket{\phi} = \left( \sum_{I, J} f_{IJ} c_{I}^\dagger c_{J}^\dagger \right)^{N/2} \ket{0}, \\
&\mathcal{P}_G=\exp \left( -\sum_{\lambda=c,s} g^{\lambda} \sum_i n_{i\uparrow \lambda} n_{i\downarrow\lambda}\right),\\
&\mathcal{P}_J=\exp \left( -\sum_{i,j}v^c_{ij} (n_{ic}-1)(n_{jc}-1)\right),
\end{align}
where $n_{i\sigma \lambda}=c^\dagger_{i\sigma \lambda} c_{i\sigma \lambda}$, $n_{i \lambda}=\sum_\sigma n_{i\sigma \lambda}$ and $N$ is the total number of electrons and localized spins. 
Indices $I$ and $J$ include site, spin and layer indices, i.e. $I=(i,\sigma_i,\lambda_i)$ and $J=(j,\sigma_j,\lambda_j)$. 
We treat $g^{c},v^{c}_{ij}$ and $f_{IJ}$ as the variational parameters. 
We fix
$g^{s}$ 
to infinity to exclude the double occupation on the localized spin layer.
We use the stochastic reconfiguration method to simultaneously optimize all the variational parameters~\cite{Sorella2001}.

\begin{figure}[htbp]
  \begin{center}
   \includegraphics[width=80mm]{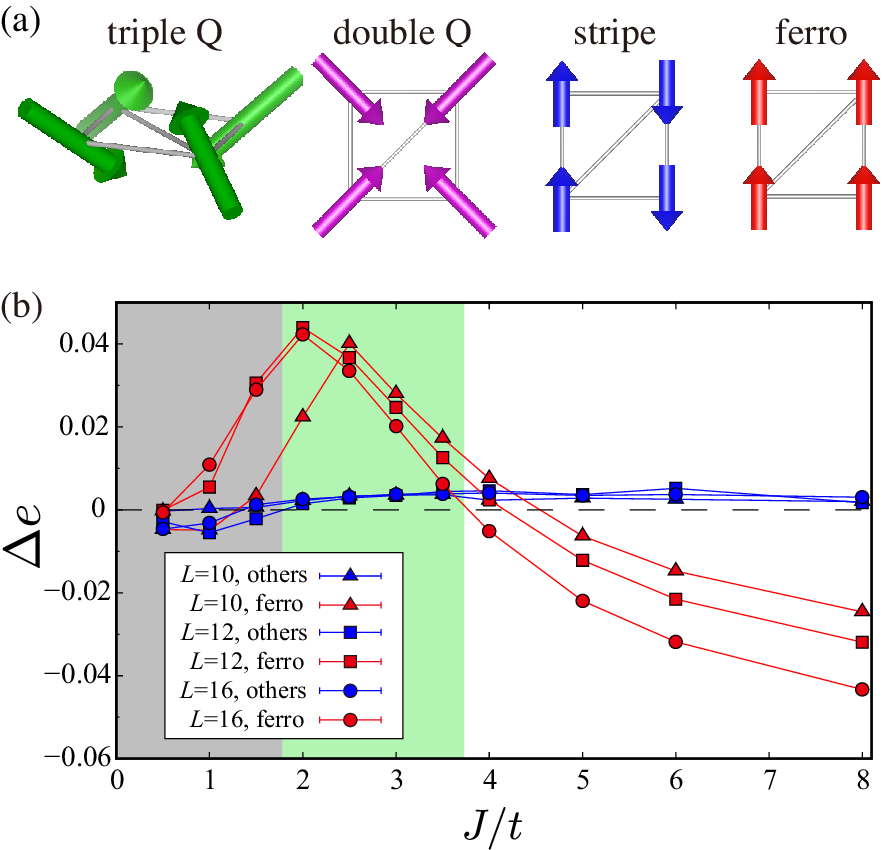}
  \end{center}
  \caption{
  (a)~Spin configuration considered as initial states for the mVMC calculations. 
  (b)$J$-dependence of the energy measured from the triple-$Q$ state, $\Delta e=(E-E_{{\rm triple-}Q})/N_s$. 
  Red symbols denote the results for the ferromagnetic state. Blue symbols represent the energy difference between the energy of the triple-$Q$ state and the lowest energy among the other magnetic states such as double-$Q$ and stripe states. 
  }
  \label{gs_af}
\end{figure} 

\begin{figure}[htbp]
  \begin{center}
   \includegraphics[width=80mm]{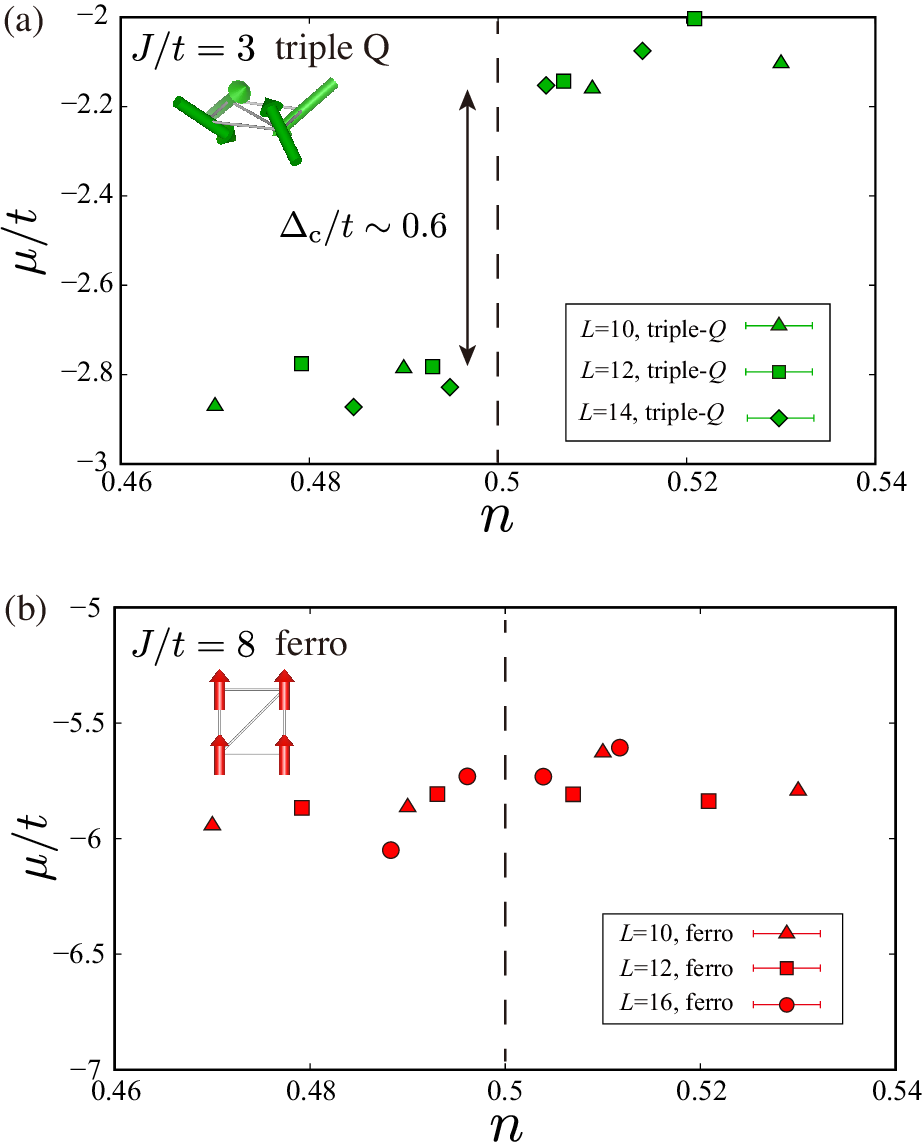}
  \end{center}
  \caption{
  Doping dependence of the chemical potential $\mu$ of (a) the triple-$Q$ state for $J/t=3$ and (b) the ferromagnetic state for $J/t=8$.
  }
  \label{mu-n}
\end{figure} 

In this study, we consider four magnetic-ordered states as candidates of the ground states, i.e.,
the triple-$Q$ magnetic ordered state, the double-$Q$ magnetic ordered state,
the stripe magnetic ordered state, and the ferromagnetic state.
Schematic pictures of these ordered states are shown in
Fig.~\ref{gs_af}(a).
To obtain the triple-$Q$ state and ferromagnetic state, the generalized pairings $f_{IJ}$ are treated as the complex numbers.
For the other states, we only optimize the real part of $f_{IJ}$.
We note that $g^{c}$ and $v^{c}_{ij}$ are treated as the real numbers.
We impose the $1\times 1$ and $2\times 2$ sublattice structures on $f_{IJ}$ 
to efficiently represent the ferromagnetic state and the other magnetic states, respectively. 
We also assume that $v^{c}_{ij}$ has the translational symmetry reflecting the lattice structure to reduce the numerical costs.

%
\section{Results}
\subsection{Ground-state phase diagram}
%
We examine the possibility of the triple-$Q$ state in
the Kondo lattice model defined in Eq.(\ref{eq:klm}).
Figure \ref{gs_af}(b) shows 
$J$-dependence of the energy measured from the triple-$Q$ 
state for $L=10,12,$ and $L=16$.
We find that the 
triple-$Q$ state becomes the ground state for $2 \leq J/t \leq 3.5$. 
We note that the triple-$Q$ state and the ferromagnetic state can always stabilize 
while the optimization of the double-$Q$ and stripe states tends to be unstable and they sometimes converge to 
other magnetic states such as the ferromagnetic state. 
Thus, in Fig.~\ref{gs_af}, we use {\it others} to describe the magnetically ordered states
other than the triple-$Q$ state and the ferromagnetic state. 

For $J/t < 2$, 
because of the severe competition among the magnetic-ordered states,
we cannot identify the ground states clearly within 
available system sizes.
In this weak coupling region, 
the long-period magnetic ordered state may be realized.
To examine the possibility of the
long-period magnetic ordered phases, 
it is necessary to extend the sublattice structure of the variational parameters. 
Since the numerical cost for larger sublattice structures is very high, 
we leave this issue for a future study.
In the strong coupling region ($J/t>3.5$),
we find that the ferromagnetic ordered phase becomes the ground state.
This is consistent with the previous studies in the classical limit~\cite{Akagi2010}.

Next, to clarify whether the obtained triple-$Q$ state is insulating or not, 
we calculate the chemical potential $\mu$ as a function of the electron density $n=N_{\rm e}/N_{\rm s}$, which is defined as
$\mu(N_{\rm e}+1)= (E(N_{\rm e}+2)-E(N_{\rm e}))/2$, where $E(N_{\rm e})$ denotes 
the total energy with the number of electrons $N_{\rm e}$. 
Fig.~\ref{mu-n}(a) shows $\mu$-$n$ plot of the triple-$Q$ state for $J/t=3$. From this plot, we evaluate the charge gap as $\Delta_{\rm c}/t\sim 0.6$.
This is quite large compared with the gap arising from the spin-orbit interaction, which is a driving force to open the charge gap in conventional topological insulators. Note that, as shown in Fig. \ref{mu-n}(b), the ferromagnetic state does not have a finite charge gap and thus is metallic.

\begin{figure}[tbp]
  \begin{center}
   \includegraphics[width=80mm]{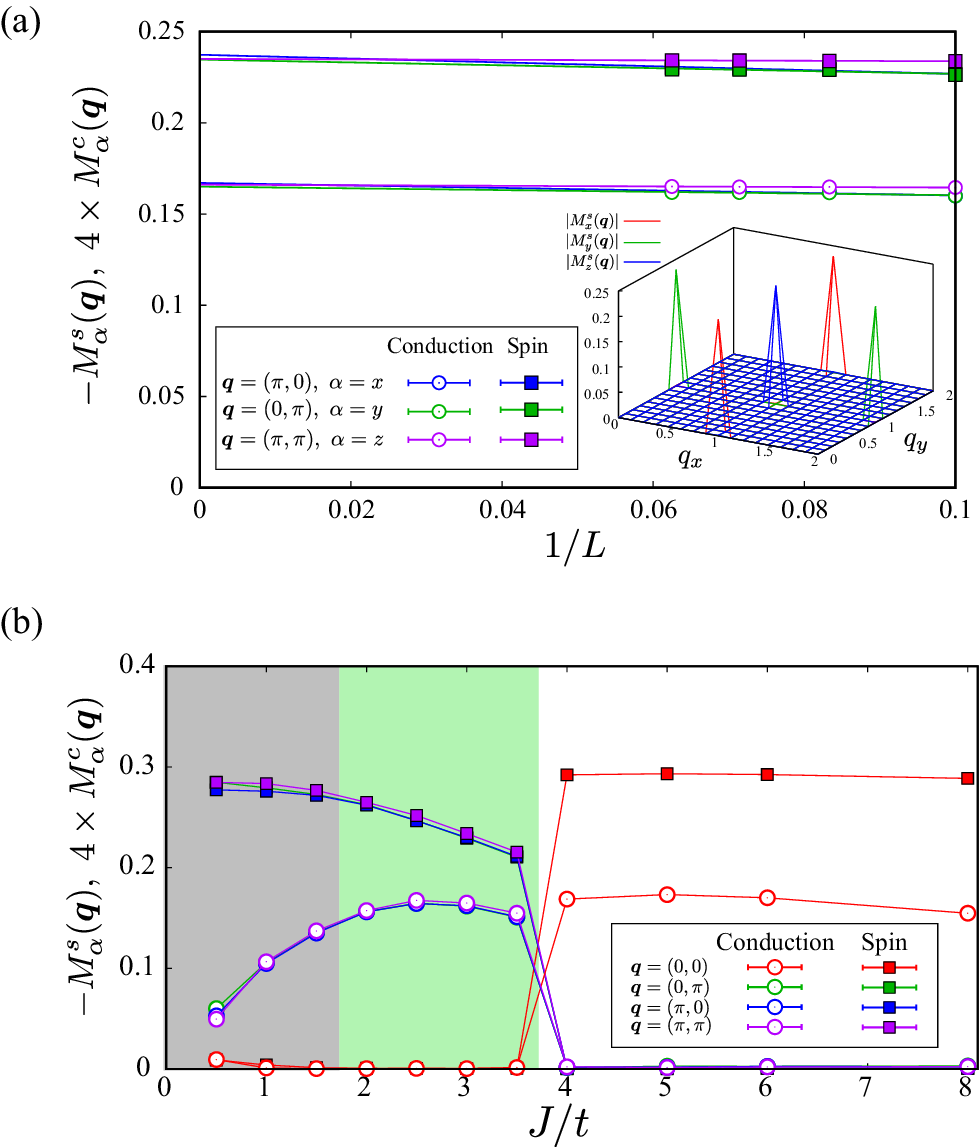}
  \end{center}
  \caption{Magnetic properties of the triple-$Q$ state and the ferromagnetic state in the triangular Kondo lattice model.
  (a)~Size-dependence of $M^\lambda_\alpha(\bm{q})$ of the ground state for $J/t=3$.
  Circles and squares represent the results for the magnetic moments $M^c_\alpha(\bm{q})$ and $M^s_\alpha(\bm{q})$, respectively.
  Colors of the symbols denote the wavenumber $\bm{q}$ and the direction of the spin operator $\alpha$.
  Thin lines shows the linear lines fitted from $M^\lambda_\alpha(\bm{q})$ for each $\bm{q}$, $\alpha$ and $\lambda$. 
  Inset shows the $\bm{q}$-dependence of the magnetic moment on the localized spin layer $|M^s_\alpha(\bm{q})|$ for $J/t=3$. Red, green, and blue lines denote $x$, $y$, and $z$ components of the magnetic moment, respectively.
  (b)~$J$-dependence of magnetic order parameters for $L=16$. 
  Squares and circles represent the results of spin and conduction layers, respectively.
  We plot $M^\lambda (\bm{q})$ of the triple-$Q$ state for $J/t <4$ and the ferromagnetic state for $J/t\geq 4$. 
  }
  \label{spin_gs}
\end{figure} 

Here, we examine magnetic properties of the ground states. 
We calculate the Fourier-transformed magnetic moment $M^\lambda_\alpha (\bm{q})=\frac{1}{N_s} \sum_{i} \braket{S^{\lambda,\alpha}_i}e^{-i\bm{q}\cdot\bm{r}_{i\lambda}}$, where $\bm{r}_{i\lambda}=(x_{i\lambda},y_{i\lambda})$ is the position vector at the $i$th site on $\lambda$ layer.
Figure \ref{spin_gs} shows $M^\lambda_\alpha(\bm{q})$ of the ground states.
To enhance the visibility, we plot $4M^c_\alpha(\bm{Q})$ and $-M^s_\alpha(\bm{Q})$. 
Note that $M^s_\alpha(\bm{Q})$ has the opposite sign of $M^c_\alpha(\bm{Q})$
since we consider the antiferromagnetic Kondo coupling.

Figure \ref{spin_gs} (a) shows the size dependence of $M^{\lambda}_\alpha (\bm{Q})$ for $J/t=3$, where $\bm{Q}$ represents an ordering wavevector. 
We also plot the $|M^{s}_\alpha (\bm{q})|$ in the momentum space 
for $L=16$ in the inset of Fig. \ref{spin_gs} (a). 
We see that $M^{s}_\alpha (\bm{q})$ 
has sharp Bragg peaks for each $\alpha$, which signal long-range magnetic order
at three different $\bm{Q}s$. 
We find that $\bm{Q}$s are $(\pi,0)$, $(0,\pi)$ and $(\pi,\pi)$ for $\alpha=x,y$ and $z$, respectively. 
As shown in Fig. \ref{spin_gs} (a), size dependence of 
$M^{s}_\alpha (\bm{q})$ and $M^{c}_\alpha (\bm{q})$ 
are small and they have finite values in the thermodynamic limit.
These results suggest that the ground state for $J/t=3$ is the triple-$Q$ state. 

Figure \ref{spin_gs} (b) shows $J$-dependence of $M^\lambda_\alpha(\bm{Q})$.
In the gray region, we plot $M^\lambda_\alpha(\bm{q})$ of the triple-$Q$ state 
for simplicity.
For $J/t <4$, $M^\lambda_\alpha(\bm{Q})$ for $\alpha=x,y$ and $z$ is finite at $\bm{Q}=(\pi,0)$, $(0,\pi)$ and $(\pi,\pi)$, respectively. This suggests that the triple-$Q$ state is indeed 
the ground state in the range of $J/t$. 
The magnetic ordered moments of the localized spins decrease by increasing $J/t$ 
due to the Kondo screening. 
The reduction of the magnetic moment, $\Delta M = 0.5- |\bm{M}^{s}|$, is about 0.1 
for $J/t=3$.
This small reduction may be related to the fact that the triple-$Q$ state has the additional $Z_{2}$ symmetry breaking of the spin scalar chirality 
as discussed in the linear-spin wave analysis~\cite{AkagiJPSJ2013}.

Before the transition to the ferromagnetic phase ($J/t\leq4$),
the magnetic ordered moments on the conduction layer show the 
non-monotonic $J/t$ dependence. 
This behavior can be attributed to the dual nature of the Kondo coupling:
It can induce both the magnetic ordered states and the singlet state.
Similar non-monotonic behavior is also found in the Kondo lattice
model on a square lattice at half filling~\cite{Capponi2001}.

Around $J/t\sim 4$, we find that there is the first transition from the 
triple-$Q$ state to the ferromagnetic state. 
Since $|\bm{M}^c|$ is much smaller than $|\bm{M}^s|$, the net magnetization of 
ferromagnetic state is finite. For $J/t > 4$, the $J$ dependence of the magnetization is small. 
This result implies that nearly all the conduction electrons contribute to the formation of the Kondo singlets. 
The reduction of the magnetization of the localized spins $\Delta M$ is about 0.2.

%
\subsection{Many-body Chern number}
%
The triple-$Q$ magnetic insulator 
is expected to have finite Chern number~\cite{Martin2008,Akagi2010}. 
To calculate the many-body Chern number,
we evaluate expectation values of the Resta's polarization 
operator with the flux insertion~\cite{Resta1998,Kang_PRL2021,Dehghani2021}.
The procedure is summarized as follows: We first introduce a flux in the hopping terms 
along the $x$ direction as 
$t_{ij} \rightarrow t_{ij} \exp (-i \theta_x (x_{ic} -x_{jc})/L)$ defined in Eq. (\ref{eq:klm}) and obtain the ground state of the Hamiltonian with the flux. 
The normalized ground-state wavefunctions is described as $\ket{\bar{\psi}(\theta_x)}$.
Then, we calculate the expectation value of the polarization operator along 
the $y$ direction, which is defined as 
\begin{eqnarray}
P_y(\theta_x) = \braket{\bar{\psi}(\theta_x) | \exp (\frac{2\pi i}{L}\sum_j n_{jc} y_{jc} ) | \bar{\psi}(\theta_x) }. \label{eq:polarization}
\end{eqnarray}
The many-body Chern number $\mathcal{C}$ is 
obtained as the slope of the argument of $P_y(\theta_x)$, namely, 
\begin{align}
  &\varphi_y(\theta_x) = \arg[P_y(\theta_x)],\\
  &\mathcal{C}= \left. \frac{\partial \varphi_y (\theta_x)}{\partial \theta_x}\right|_{\theta_x \rightarrow 0}.
\end{align}
We note that the expectation value of the Resta's polarization operator in Eq. (\ref{eq:polarization})
can be easily calculated by the VMC method since it only
includes the real-space diagonal operator $n_{jc}y_{jc}$.
We also note that it is pointed out that the thermodynamic limit of the Resta's 
polarization formula may need special care in more than two dimensions~\cite{Watanabe2018}. 
 
\begin{figure}[tbp]
  \begin{center}
   \includegraphics[width=80mm]{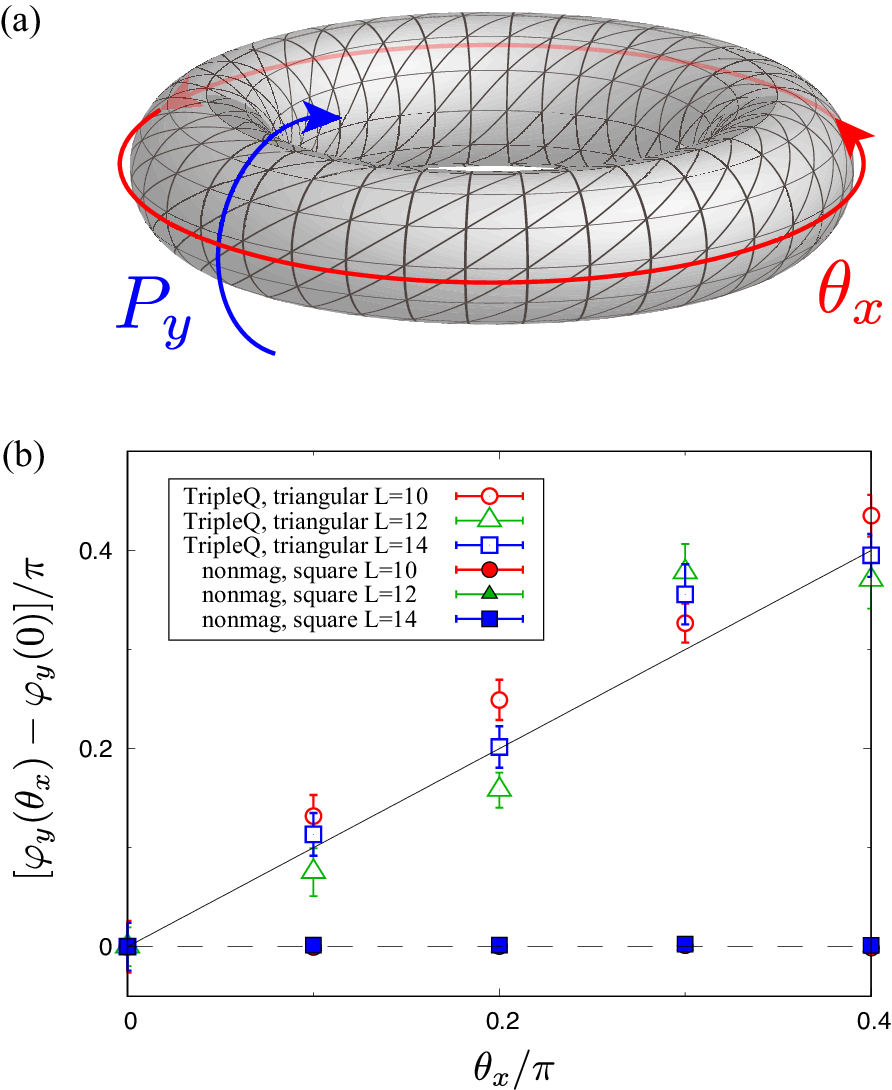}
  \end{center}
  \caption{
(a)~Schematic figure of the polarization operator along the $y$ direction ($P_y$) and the flux $\theta_x$ along the $x$ direction on the torus.
  (b)~Flux-dependence of the argument of the polarization $\varphi_y(\theta_x)$ in the 
  triple-$Q$ state at $J/t=3$ (open symbols). 
  For reference, we calculate $\varphi_y(\theta_x)$ for the non-magnetic Kondo insulator
  in the Kondo lattice model on the square lattice at half filling and $J/t=3$ (closed symbols). See Eq. (\ref{eq:klm}) for the definition of the Kondo lattice model.
  Note that the variational parameters in the pair product part 
  $f_{IJ}$ are treated as complex numbers in these calculations. 
  As a guide for the eye, we 
  plot the thin line for $\varphi_y(\theta_x) = \theta_x+\varphi_y(0)$ and the dashed line for $\varphi_y(\theta_x) = \varphi_y(0)$.
  }
  \label{resta}
\end{figure} 

Figure \ref{resta} shows $\theta_x$-dependence of $\varphi_y(\theta_x)$ 
in the triple-$Q$ state for $J/t=3$. 
In this state, we find that
$\varphi(\theta_x)$ 
linearly increases as a function of $\theta_{x}$ and its slope is 
evaluated as $1.007(74)$ for $L=14$ by the linear regression.
This result indicates that the triple-$Q$ state is 
the many-body Chern insulator with $\mathcal{C}=1$.
For comparison, we also calculate $\varphi(\theta_x)$ for a nonmagnetic Kondo insulator in the square lattice at half filling~\cite{Assaad1999,Capponi2001}, which is expected to be
a topologically trivial insulator.
As a result, we find that the slope of  $\varphi(\theta_x)$ is almost zero
in the nonmagnetic Kondo insulator, which indicates that this state
is a trivial insulator.

%
\section{Summary and Discussion}
%
In summary, we have performed the mVMC calculations 
for the Kondo lattice model on the triangular lattice at quarter filling. 
We find that the triple-$Q$ magnetic ordered phase becomes the ground state in the intermediate coupling region. 
By explicitly calculating the many-body Chern number, 
we show that the triple-$Q$ state is indeed magnetic Chern insulators
with $\mathcal{C}=1$.

Here, we discuss the stability of the triple-$Q$ 
state in the Kondo lattice model. 
In the classical limit and the spin wave analysis\cite{Akagi2010, Akagi2012, AkagiJPSJ2013}, the 
triple-$Q$ magnetic order becomes the ground states for $2.8<J/t<18$, which is quite robust compared with our results~\cite{DefJ}. 
Our results indicate the quantum effects significantly destabilize the triple-$Q$ state. This destabilization may be attributed to the absence of the effective biquadratic term
in the Kondo lattice model.
In the classical limit at quarter filling, 
it has been shown that the effective positive biquadratic terms derived by a perturbation theory
play an important role in inducing the noncoplanar spin configuration~\cite{Akagi2012}.
It is also discussed that the positive biquadratic term is significant
in understanding the triple-$Q$ state in itinerant magnets~\cite{Takagi2018,park2023tetrahedral,takagi2023spontaneous}.
However, this term does not exist in the Kondo lattice model with spin 1/2 since
it is equivalent to the ferromagnetic Heisenberg interaction with an additional constant term.
This ferromagnetic Heisenberg interaction stabilizes the collinear ferromagnetic state and does not induce the noncoplanar triple-$Q$ state.
Nevertheless, we show that the triple-$Q$ state still survives  
in the Kondo lattice model.
This indicates that itinerant magnets including spin 1/2 localized spins
can have the triple-$Q$ state. 
Furthermore, the triple-$Q$ state with a large charge gap is expected to be more thermally stable than other Chern insulators found in previous studies, most of which are realized below about 20 K \cite{deng2020quantum, okazaki2022quantum,takagi2023spontaneous}.
Thus, our results are expected to stimulate 
the exploration of the triple-$Q$ state at high temperatures
and the resultant magnetic Chern insulator 
in a wide range of correlated electron systems.
One of the possible candidates is a van der Walls heterostructure recently found in the experiment \cite{vavno2021, takagi2023spontaneous}.

Lastly, we discuss how to further stabilize the triple-$Q$ state. 
One promising way to stabilize the triple-$Q$ state is
adding the onsite Coulomb interaction to the conduction layer. 
In the triangular Hubbard model at quarter filling, 
the mean-field results suggest that the noncoplanar state becomes the ground state in the intermediate coupling region~\cite{li2010metalinsulator}.
The recent mean-field study also shows that 
skyrmion crystals with triple-$Q$ spin structures emerge around quarter filling\cite{Kobayashi2022}.
Analysis of such an extended Kondo lattice model on the triangular lattice is an
intriguing issue but is also left for future study.

%
\begin{acknowledgments}
%
We use {\tt VESTA} to make Fig. \ref{klm} and Fig. \ref{gs_af} (a) \cite{momma2011vesta}.
We acknowledge Y. Akagi for fruitful discussions.
We also thank the Supercomputer Center, the Institute for Solid State Physics, the University of Tokyo 
and Oakbridge-CX in the Information Technology Center, the University of Tokyo for their facilities.
This work was financially supported by Grant-in-Aid for Scientific Research 
(Nos. JP16H06345, JP19K03739, JP19K14645, JP23H03818 and JP23K13055) from Ministry of Education, Culture, Sports, Science and Technology, Japan.
TM was supported by Building of Consortia for the Development of 
Human Resources in Science and Technology from the MEXT of Japan. 
\end{acknowledgments}
\bibliography{reference}

\begin{thebibliography}{62}
\bibitem [{Def()}]{DefJ}%
  \BibitemOpen
  \href@noop {} {}\bibinfo {howpublished} {Note that the exchange interactions
  in the classical Kondo-lattice model $J_{\rm cl}$ is four times smaller than
  the exchange interactions defined in Eq. (\ref{eq:klm}), i.e., $J_{\rm
  cl}=J/4$.}\BibitemShut {Stop}%
\end{thebibliography}

\end{document}